# Fabrication of Conducting Si Nanowire Arrays

E. Johnston-Halperin,[1] R. A. Beckman,[1] N. A. Melosh,[2] Y. Luo, J. E. Green,

and J. R. Heath[3]

*Division of Chemistry and Chemical Engineering, MC 127-72 California Institute of*

*Technology, Pasadena, CA 91125*

**Abstract**

The recent development of the superlattice nanowire pattern transfer (SNAP) technique allows for the fabrication of arrays of nanowires at a diameter, pitch, aspect ratio, and regularity beyond competing approaches. Here, we report the fabrication of conducting Si nanowire arrays with wire widths and pitches of 10-20 nm and 40-50 nm, respectively, and resistivity values comparable to the bulk through the selection of appropriate silicon-on-insulator substrates, careful reactive-ion etching, and spin-on glass doping. These results promise the realization of interesting nano-electronic circuits and devices, including chemical and biological sensors, nano-scale mosaics for molecular electronics, and ultra-dense field-effect transistor (FET) arrays.

[1] These authors contributed equally to this work.

[2] Current Address: Stanford University, Department of Materials Science & Engineering

[3] Corresponding author: heath@caltech.edu

As complementary metal oxide semiconductor (CMOS) technology is scaled to the dimensions of a few nanometers, a key challenge is the manufacture of conducting silicon circuit elements at small dimensions and narrow pitch. Recent reports have partially addressed these challenges through, for example, the fabrication of CMOS field-effect transistors (FETs) with gate lengths of 10-20 nm,[1] or of conducting semiconductor nanowires.[2] These studies constitute single- or few-device demonstrations of scaling feature size, and neglect considerations such as feature pitch, device-to-device reproducibility, and manufacturability. The full realization of nano-scale electronics will require not just the making small devices, but making small devices in high density and with high reproducibility. Overcoming these additional challenges would open the door to a variety of applications ranging from highly parallel chemical sensor arrays[3] with implications for molecular biology[4] to nano-mosaics for molecular based computing.[5]

To that end, we have developed the superlattice nanowire pattern transfer (SNAP) technique for fabricating arrays of aligned, high aspect ratio, metal and semiconducting nanowires.[6] Here, we employ SNAP to fabricate high-density conducting Si nanowire (NW) arrays. We find that NWs with widths below 50 nm are critically sensitive to defects introduced by standard processing techniques. However, through a combination of careful substrate selection, spin-on glass doping, and low-energy fluorine-chemistry based reactive ion etching (RIE), we demonstrate that Si NWs at diameters of 10-20 nm, pitches of 40-50 nm, and lengths in excess of 1 millimeter can be fabricated with controllable, bulk-like conductivity characteristics. Furthermore, the SNAP technique can produce NWs with widths and pitches as small as 8 nm and 16 nm, respectively, presenting no apparent limitations to the fabrication of NW crossbar circuits at densities of at least $10^{12}$ junctions/cm$^2$.

Samples were fabricated using a variation of the SNAP technique that has been described previously.[6] Briefly, the GaAs layers in a freshly cleaved GaAs/Al$_{0.5}$Ga$_{0.5}$As superlattice are selectively wet etched to form a comb structure (Figure. 1A). Platinum is then deposited along the ridges of the comb,[7] and the resulting Pt wires are gently pressed into a thin epoxy layer spun onto a silicon-on-insulator (SOI) substrate. The superlattice templates are removed by wet-chemical etching, and an oxygen-plasma etch is utilized to remove residual epoxy between the wires. The Pt wires serve a shadow mask in a final dry-etch used to transfer the NW pattern into the SOI substrate. The result is an array of Si NWs on an insulating oxide (Fig. 1B).

In order to maximize NW conductivity, we investigate the relative importance of defects native to our SOI substrates as compared to defects introduced in the SNAP fabrication process. A p-doped Si epilayer grown via molecular beam epitaxy (MBE) represented our 'gold standard' for fabricating high quality NWs; this substrate was used to test the reactive ion etch (RIE) transfer of the NW pattern from Pt to Si. Those wires then served as a metric against which commercially available 4" and industry standard 8" SOI wafers were compared.[8] Additonally, the 4" SOI substrates were used to compare ion implantation doping versus spin-on doping methods. These three factors – choice of starting material, etching chemistry, and doping method – are all essential for obtaining high quality NWs.

The electrical resistivity of the Si NWs provides our figure of merit for nanowire quality. As a consequence, Si NW arrays were sectioned into multiple regions of differing lengths, from 5 – 25 µm long,[9] and each section was contacted with four metal leads (Ti/Al/Pt 10 nm/150 nm/20 nm) then subsequently annealed to promote an Ohmic contact (Ar; 450° C; 5 minutes). This allows for the measurement of two sets of wires per region, and for cross-conductance measurements between each set. Individual wires are 10 - 20 nm in width, depending on process conditions, and 2-4 wires are addressed by each contact (Fig. 2A, lower right inset).

We employed a Unaxis SLR parallel plate RIE system to maximize our control over the critical Si etch step. This tool employs a high frequency power supply (40 MHz versus the more common 13.56 MHz) that generates a stable plasma at DC bias as low as 10 V. An etch-gas mix of $CF_4$, $H_2$, and He provided protection for the Si sidewalls while attacking the exposed Si substrate.[10] This etch recipe achieved high-fidelity pattern transfer with vertical Si sidewalls and no noticeable undercut as observed with scanning electron microscopy (SEM).

Figure 2A (filled circle) shows a current-voltage trace from a 7 μm long set of nanowires fabricated from the MBE substrate (30 nm B:Si on i-Si, $p = 1 \times 10^{19}$ cm$^{-3}$); the linearity of the curve confirms the Ohmic nature of the electrical contacts. A histogram representing several such measurements normalized by the bulk-scaled resistance, $R_0$, reveals bulk-like conductivity characteristics for these wires (Fig. 2B, top).[11] The good morphological properties of these NWs apparently correspond to good electronic properties, confirming that our selection of etch recipe has prevented significant damage to the NWs.

In contrast, when Si NWs were fabricated from an ion-implantation doped 4" SOI wafer (B:Si, 25 nm Si on 150 nm of oxide, $p = 3 \times 10^{19}$ cm$^{-3}$), the corresponding $R/R_o$ histogram is centered near $10^4$ (not shown), indicating that the electrical properties of those NWs are severely degraded. A probable defect source is lattice damage caused by the ion-implantation process that is imperfectly healed by post-doping thermal annealing. Using the ion-doped substrates, we find that we can use SNAP to fabricate highly conductive Si NWs down to 50 nm in width, but 10-25 nm wide wires are routinely poor conductors. This can be explained if one assumes both that a single defect is able to block all conduction pathways in the narrow NWs, and that there are a sufficient number of such defects to guarantee at least one per NW. We therefore explored spin-on doping as a batch-processing compatible doping alternative. This method utilizes a high-temperature anneal to diffuse dopant atoms into the Si epilayer. A brief summary of our optimized process follows.

We used Emulsitone Phosphorosilicafilm or Emulsitone Borofilm[12] 100 (each at dopant concentrations of $5\times10^{20}$ cm$^{-3}$) diluted 1:10 in methanol as our n-type and p-type dopant sources, respectively. The solutions were generously applied to a clean substrate surface, the wafer was spun at 4000 RPM for 30 seconds, and heated at 200 C to remove excess solvent. The substrate was then annealed under Ar, with the anneal conditions determined from a look-up table.[13] An n-type spin-on dopant matrix was removed with a buffered oxide etch (BOE) (6:1, NH$_4$F to HF) while a p-type matrix was removed using hot 2:1, H$_2$SO$_4$ : H$_2$O$_2$, followed by BOE.

To test the efficacy of this doping process, we fabricated SNAP patterned Si NWs using the spin-on doping method on 4" SOI wafers (25 nm thick Si on 150 nm of oxide, sections 3μm long, p = $5 \times 10^{18}$ cm$^{-3}$). A slight non-linearity in the current response at low voltages (Fig. 2A, filled diamond) is recorded, although measurements of the length-dependent resistance exhibit linear scaling,[14] suggesting that the observed resistance values are dominated by the native wire conductivity. For these NWs, as for all NWs fabricated from SOI substrates, the resistance between sets of NWs is found to be > 1 TΩ. Resistance histograms of these NWs (Fig. 2B, middle) reveal that, while their conductivity is worse than our gold standard by a factor of 10, it is significantly better than the ion-implantation doped NWs, confirming ion-bombardment damage as the leading cause of suppressed conductivity in the latter NWs.

A third potential source of defects is the method used to prepare the undoped SOI wafers. The two SOI wafers are produced via oxygen-ion bombardment and thermal annealing, which can result in conductivity-limiting defects in a fashion similar to that described for ion-doping. Our SOI substrates have defect densities of < 0.1 cm$^{-2}$ for the 8" (industry standard) wafer and 0.23 cm$^{-2}$ for the 4" wafer. We also found substantial leakage current through the insulating oxide of the 4" SOI wafer, but no such problems with the 8" wafer, implying that the 4" wafer was of substantially lower quality. Figure 2A (filled square) and the bottom histogram in Fig. 2B show a representative current-voltage trace and normalized resistance values, respectively, for SNAP Si NW arrays fabricated on an 8" SOI wafer (30.8 nm thick Si on 145nm of oxide, sections 3μm long, p = 5 × 10$^{19}$ cm$^{-3}$). The resistance histogram reveals bulk scaling of the resistivity, validating our choices of substrate, doping technique, and etch recipe. We also prepare n-doped NWs on the 8" SOI substrate (n = 1 × 10$^{20}$ cm$^{-3}$). A representative current-voltage scan of 3 μm long wires is shown in Fig. 2A (filled triangle, dashed line), confirming that our fabrication process is compatible with both p- and n-type dopants. Normalized resistance data for the n-type wires are consistent with that observed for p-type wires.

Finally, we assess the voltage response of these NWs by constructing a nano-FET structure. For this measurement 7 nm of Al$_2$O$_3$ was deposited over a NW array fabricated from the 4" SOI wafer (7.5 μm long wires, p = 5 × 10$^{18}$ cm$^{-3}$) and a finger gate 250 nm wide was deposited across the nanowires. Figure 3 shows a representative data set where we plot the current response as a function of gate voltage for various values of the source-drain voltage. The data reveal that we are able to modulate the conductivity by three orders of magnitude with the relatively modest gate voltage of +/- 1 V. This voltage sensitivity is comparable to that of Si nanowires grown using vapor-liquid-solid growth techniques[15] that have been shown to function as chemical and biological sensors,[16] suggesting that these NWs will also be suitable for that purpose. In addition, appropriate patterning of the gate oxide should enable the construction of a binary-tree multiplexer that can bridge length scales,[17] enabling us to address each wire within SNAP arrays with pitches as small as the current fabrication limit (16 nm) and beyond.

These results highlight the importance of choosing appropriate substrates and processing methods for fabricating high-density nanoscale circuits. We show that high quality nanostructures may be fabricated from commercially available and relatively inexpensive substrates using processing techniques derived from current practice. This result should provide a useful complement to existing techniques of nanowire synthesis that, while capable of producing high quality wires in either single-wire or batch fabrication,[18] pose significant challenges in terms of scalability, assembly, and manufacturability.

This work was supported by the DARPA, the MARCO Materials Structures and Devices Focus Center, and the Semiconductor Research Corporation. We acknowledge Dr. Kris Beverly and Akram Boukai for fabrication assistance.

[10] The full etch recipe is as follows: $CF_4/He/H_2$ at 20 sccm/30 sccm/2.5 sccm respectively and an overall pressure of 5 mT. The RF power was 40 W and the etch time was determined using an end-point detector (interferometer). A full explanation of the mechanism of this etch can be found in the following references: S. Wolf and R. N. Tauber, "Silicon processing for the VLSI era" (Lattice Press, Sunset Beach, California, 2000); "Handbook of advanced plasma processing techniques," R. J. Shul and S. J. Pearton, eds., (Springer-Verlag, Berlin Heidlber, 2000).

[11] This value is calculated using the dimensions of the nanowires as measured via SEM and using the resistivity measured for the bulk substrate assuming 3 NWs are contacted for each electrode. Fluctuations due to the varying numbers of wires contacted are not significant on the scale of the fluctuations between substrates.

[12] Emulsitone Company, Whippany, New Jersey.

[13] The look-up table was generated by measuring the 4-point resistivity of a series of doped samples annealed at various temperatures and times. Temperatures ranged from 800° C to 1000° C and times from 1 – 8 minutes.

[14] Resistance values were measured for wires from 2.5 μm to 22.5 μm in length.

[15] A. M. Morales and C. M. Lieber, *Science* **279**, 208 (1998);

[16] Y. Cui, Q. Wei, H. Park, and C. M. Lieber, *Science* **293**, 1289 (2001).

[17] J. R. Heath and M. A. Ratner, *Physics Today,* pg. 43, May 2003.

[18] T. Ono, H. Saitoh, and M. Esashi, *Appl. Phys. Lett.* **70**, 1852 (1997); A. Tao, F. Kim, C. Hess, J. Goldberger, R. He, Y. Sun, Y. Xia, and P. Yang, *Nano. Lett.* **3**, 1229 (2003); Y. Yin, B. Gates, and Y Xia, *Adv. Mat.* **12**, 1426 (2000); L. S. Li and A. P. Alivisatos, *Adv. Mat*. **15**, 408 (2003).

**Figure Captions**

**Figure 1.** Steps of the SNAP process utilized to fabricate the Si nanowire arrays are illustrated here. **A.** The GaAs/ Al0.5Ga0.5As master, with the GaAs partially etched to produce the SNAP template. **B.** 128 12 nm wide Si wires generated by using the Pt wires, formed onto and deposited from the template, as an etch mask. The inset reveals the structural fidelity of the nanowires at higher resolution. The scale bar is 150 nm.

**Figure 2. A.** IV measurements of four different nanowire samples: (square) 8" SOI, p = $5 \times 10^{19}$ cm$^{-3}$ (10nm×31nm×3 µm), (triangle) 8" SOI, n = $1 \times 10^{20}$ cm$^{-3}$ (10nm×31nm×3 µm), (circle) MBE, p = $1 \times 10^{19}$ cm$^{-3}$ (10nm×30nm×7 µm), (diamond) 4" SOI, p = $5 \times 10^{18}$ cm$^{-3}$ (10 nm×25nm×2.5µm). (upper left inset) The same data plotted on a semi-log scale. (lower right inset) SEM of Ti/Al/Pt electrodes contacting 3-4 Si nanowires each, scale bar is 500 nm. **B.** Statistical distribution of nanowire resistance normalized by the bulk-scaled resistance ($R_0$) for wires fabricated from MBE substrates (p = $1 \times 10^{19}$ cm$^{-3}$), 4" SOI (p = $5 \times 10^{18}$ cm$^{-3}$), and 8" SOI (p = $5 \times 10^{19}$ cm$^{-3}$), respectively. Horizontal axis is log-scale and bin size is 10.

**Figure 3.** Current response versus gate voltage, $V_G$, of a 7.5 µm long section of NWs fabricated from the 4" SOI wafer (cross section of 15 nm × 25 nm, p = $5 \times 10^{18}$ cm$^{-3}$) for a variety of different source-drain voltages, $V_{SD}$. The asymmetry in source-drain bias is due to a difference in Schottky-barrier height between the two contacts.

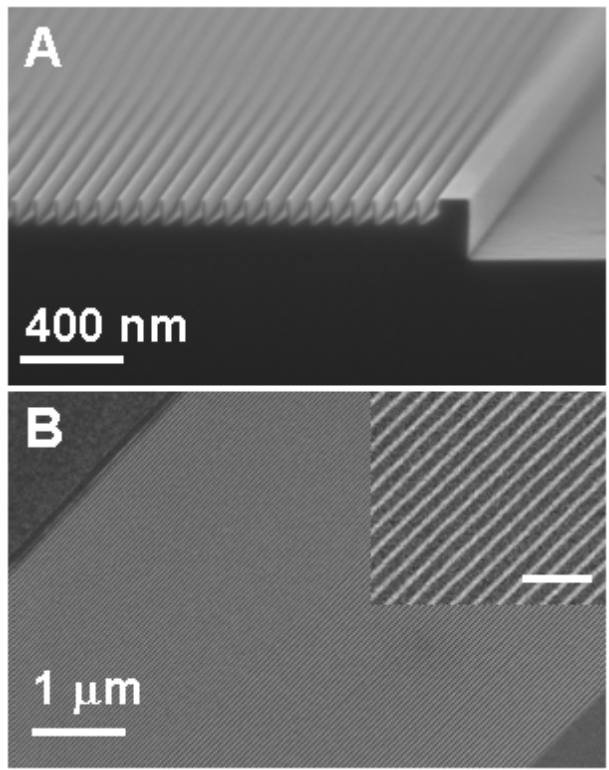

Figure 1, Johnston-Halperin, *et al.*, APL

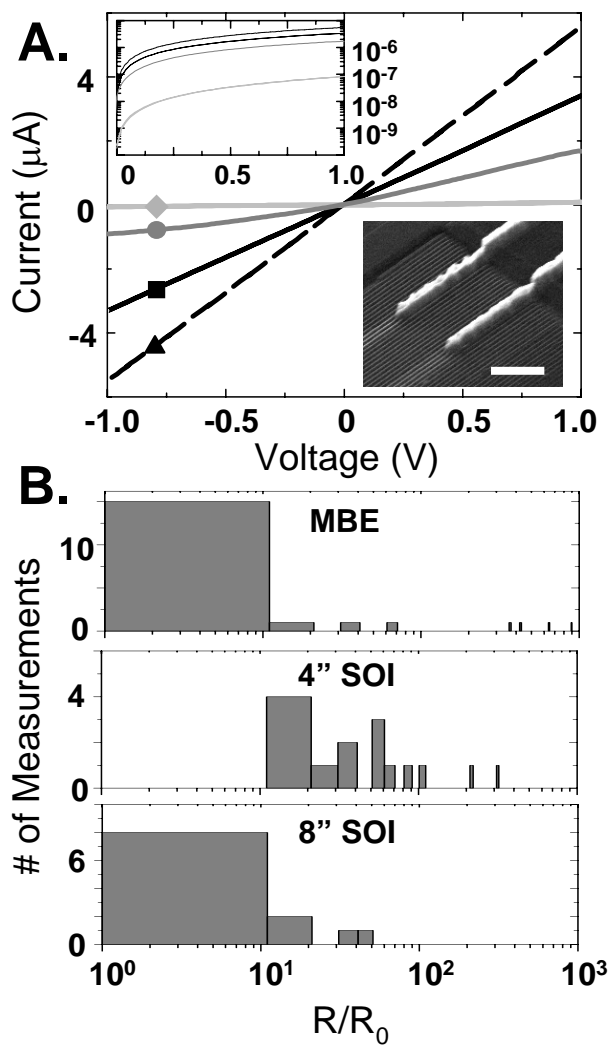

Figure 2, Johnston-Halperin, *et al.*, APL

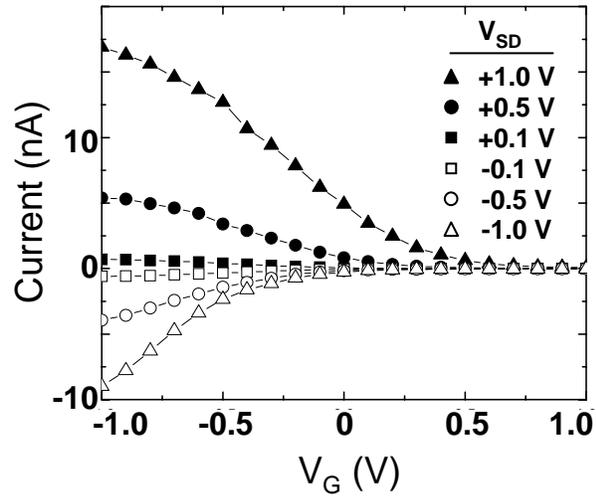

Figure 3, Johnston-Halperin, *et al.*, APL